\documentclass[reprint,aps,pra,twocolumn,superscriptaddress,floatfix,longbibliography]{revtex4-1}

\usepackage[dvips]{graphicx}
\usepackage{latexsym}
\usepackage{amsmath}
\usepackage{amsthm}
\usepackage{amsfonts}
\usepackage{amssymb}
\usepackage{mathptmx}
\usepackage{subfigure}
\usepackage{mathrsfs}

\usepackage{color}
\definecolor{darkblue}{rgb}{0,0.02,0.45}
\usepackage[colorlinks=true,citecolor=darkblue]{hyperref}

\usepackage[all]{hypcap}
\usepackage{gensymb}

\definecolor{cream}{RGB}{222,217,201}     

\begin{document}
%
%This manuscript has been authored by UT-Battelle,
%LLC under Contract No. DE-AC05-00OR22725 with
%the U.S. Department of Energy. The United States
%Government retains and the publisher, by accepting
%the article for publication, acknowledges that the
%United States Government retains a non-exclusive, paidup,
%irrevocable, world-wide license to publish or reproduce
%the published form of this manuscript, or
%allow others to do so, for United States Government
%purposes. The Department of Energy will
%provide public access to these results of federally
%sponsored research in accordance with the DOE
%Public Access Plan(http://energy.gov/downloads/doepublic-access-plan).
\clearpage

\title{Successive dielectric anomalies and magnetoelectric coupling in honeycomb Fe$_4$Nb$_2$O$_9$}

\author{Lei Ding}
\email{dingl@ornl.gov}\affiliation{Neutron Scattering Division, Oak Ridge National Laboratory, Oak Ridge, Tennessee 37831, USA }

\author{Yan Wu}
\affiliation{Neutron Scattering Division, Oak Ridge National Laboratory, Oak Ridge, Tennessee 37831, USA }

\author{Minseong Lee}
\author{Eun Sang Choi}
\affiliation{National High Magnetic Field Laboratory, Florida State University, Tallahassee, FL 32306-4005, USA}
\author{Ryan~Sinclair}
\affiliation{Department of Physics and Astronomy, University of Tennessee, Knoxville, Tennessee 37996-1200, USA}
\author{Haidong~Zhou}
\affiliation{Department of Physics and Astronomy, University of Tennessee, Knoxville, Tennessee 37996-1200, USA}
\affiliation{National High Magnetic Field Laboratory, Florida State University, Tallahassee, FL 32306-4005, USA}

\author{Bryan C. Chakoumakos}
\author{Huibo Cao}
\email{caoh@ornl.gov}\affiliation{Neutron Scattering Division, Oak Ridge National Laboratory, Oak Ridge, Tennessee 37831, USA }

\date{January 21, 2020}

\begin{abstract}

By combining single crystal x-ray and neutron diffraction, and the magnetodielectric measurements on single crystal Fe$_4$Nb$_2$O$_9$, we present the magnetic structure and the symmetry-allowed magnetoelectric coupling in Fe$_4$Nb$_2$O$_9$. It undergoes an antiferromagnetic transition at T$_N$=93 K, followed by a displacive transition at T$_S$=70 K. The temperature-dependent dielectric constant of Fe$_4$Nb$_2$O$_9$ is strongly anisotropic with the first anomaly at 93 K due to the exchange striction as a result of the long range spin order, and the second one at 70 K emanating from the structural phase transition primarily driven by the O atomic displacements. Magnetic-field induced magnetoelectric coupling was observed in single crystal Fe$_4$Nb$_2$O$_9$ and is compatible with the solved magnetic structure that is characteristic of antiferromagnetically arranged ferromagnetic chains in the honeycomb plane. We propose that such magnetic symmetry should be immune to external magnetic fields to some extent favored by the freedom of rotation of moments in the honeycomb plane, laying out a promising system to control the magnetoelectric properties by magnetic fields.
\end{abstract}


\maketitle
\section{INTRODUCTION}
\begin{figure}
\centering
\includegraphics[width=1\linewidth]{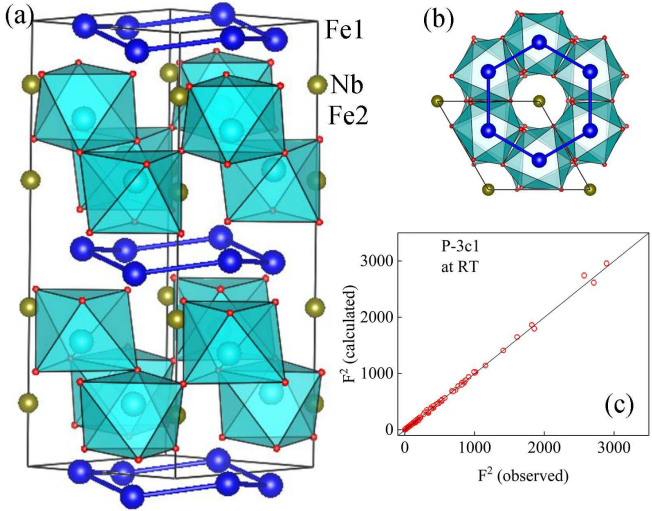}
\caption{(Color Online) (a) Schematic drawing of the crystal structure of Fe$_4$Nb$_2$O$_9$ at room temperature viewing the [110] direction. (b) Buckled honeycomb layer is formed by Fe1 atoms projected along the $c$ axis.(c) Observed and calculated single crystal x-ray diffraction data at room temperature}\label{fig:1}
\end{figure}

The linear magnetoelectric (ME) effect, which defines the control of electric polarization by an external magnetic field or the magnetization by an electric field, \citep{schmid1994, rivera2009} has attracted considerable interest because of its importance in understanding the novel types of ferroic order \citep{spaldin2008, schmid2008, fiebig2005, tokura2014} in solids and potential applications in spintronics\citep{wang2003, spaldin2005, cheong2007, eerenstein2006, tokura2014}. For the presence of the linear ME effect in a specific material, the symmetry requires the breaking of both spatial inversion and time reversal.\citep{fiebig2005} In this regards, the linear ME effect can be predicted by symmetry considerations given the magnetic symmetry of a system is clearly known.\citep{schmid1994, ding2016} 
A family of materials A$_4$Nb$_2$O$_9$ (A= Mn, Co) belonging to the corundum-type structure has been found in their powder forms to show magnetoelectric effect for several decades.\citep{Fischer1972} They crystallize with the $\alpha$-Al$_2$O$_3$-type trigonal crystal structure with the space group $P\bar{3}c1$ \citep{bertaut1961}(see Fig.\ref{fig:1}(a-b)) and can be regarded as a derivative of Cr$_2$O$_3$, one of the first predicted and discovered room temperature ME materials.\cite{Dzyaloshinskii1959, astrov1960, mcgurie1956, fiebig1994, kimura2013} The magnetic structures of both Mn and Co cases were first determined by Bertaut $et~al.$ in Ref. \cite{bertaut1961} where both of them can be simply described by an antiferromagnetically coupled collinear ferromagnetic Co$^{2+}$ chains with the moments along the $c$-axis. Originally, this symmetry allows a linear ME effect and seems to be consistent with the observed linear ME response in the powder samples \citep{Fischer1972, fang2014, Fang2015}. 

Recently, this family has attracted some renewed interest because of the successful growth of Co$_4$Nb$_2$O$_9$ single crystals and the finding of manipulations of its ME effect via external magnetic fields.\citep{khanh2017} Single crystal neutron diffraction experiments have revealed a different magnetic structure with all moments in the ${ab}$ plane.\citep{Ding2020} This allows different magnetoelectric effect terms as observed in the electric polarization measurement on a single crystal.\citep{khanh2016, Ding2020, deng2018} Fe$_4$Nb$_2$O$_9$ emerging as a new member in this family has been recently reported to show linear magnetoelectric effect accompanied by a long range antiferromagnetic ordering below 90 K and an anomaly at 77 K in the dielectric permittivity curve on a powder sample. \citep{Maignan2018} It was later shown in Ref. \citep{Jana2019} that the magnetic structure is described by collinearly arranged moments in the $ab$ plane which can be assigned in either $C2/c'$ or $C2'/c$ magnetic symmetry through a powder neutron diffraction experiment. To elucidate the exact magnetic symmetry that is vital in explaining the magnetoelectric coupling, single crystal neutron diffraction is required.   

In contrast to the previous works on a powder Fe$_4$Nb$_2$O$_9$ sample, in this work, the availability of the sizable single crystal allowed us to study the magnetoelectric coupling in different crystal directions and to unveil the magnetic ground state using the single crystal neutron diffraction technique. We show the presence of anisotropic dielectric constant and linear magnetoelectric response around the N\'eel temperature in Fe$_4$Nb$_2$O$_9$, in good accordance with the magnetic symmetry $C2/c'$ that allows a linear ME effect with both diagonal and off-diagonal components.

\section{RESULTS}
\subsection{CRYSTAL STRUCTURE AT ROOM TEMPERATURE}

\begin{table} 
\caption{The refined structural parameters of Fe$_4$Nb$_2$O$_9$ based on both single crystal x-ray (at room temperature, upper line) and neutron diffraction (at 120 K, lower line) data. (Space group: $P\bar{3}c1$, $a$=$b$=5.2118(3) \AA, $c$=14.2037(8)(4) \AA, $\chi^2$(x-ray)=0.94, $\chi^2$(neutron)=13.8, $B_{iso}$ in 1/(8$\pi^2$)\AA$^2$.)}\label{str}
\begin{tabular}{ccccccc}
\hline
\hline
atom & $x$& $y$ &  $z$  & $B_{iso}$ \\
\hline
Nb1 & 0 & 0 &  0.14251(6) & 0.56(1) \\
    & & &           0.14266(2) & 0.67(8) \\
Fe1 & 1/3 & 2/3 &  0.19325(6) & 0.77(2) \\
    & & &           0.1933(2) & 0.82(8) \\
Fe2 & 1/3 & 2/3 & 0.98522(6) & 0.72(2)\\
    & & &           0.9860(1) & 0.61(7) \\
O1  & 0.6625(6) & 0.9767(7) & 0.0847(2) & 0.85(45)\\
    &  0.6609(3)  &0.9763(5) &  0.0852(1)     & 0.72(7) \\
O2  &  0.2837(9) & 0 &  1/4 & 0.75(6)\\
    &  0.2853(6)&  &           & 0.64(8) \\
\hline
\hline
\end{tabular}
\end{table}
The availability of single crystal Fe$_4$Nb$_2$O$_9$ enabled us to carefully examine its crystal structure using both single crystal neutron and x-ray diffraction techniques, hence providing more precise crystal structural parameters. Room temperature x-ray single crystal diffraction data show that Fe$_4$Nb$_2$O$_9$ crystallizes with the space group $P\bar{3}c1$, in good agreement with the previous reports \citep{Maignan2018,Jana2019}. The refinement result based on the x-ray diffraction data with a comparison between the observed and the calculated squared structure factor is shown in Fig. \ref{fig:1}(c). Single crystal neutron diffraction at 120 K confirms the space group $P\bar{3}c1$ and reveals a good quality of the single crystal. The refined structural parameters were summarized in Table \ref{str}. As shown in Fig. \ref{fig:1}, the crystal structure is characterized by an alternate stacking of slightly buckled Fe1O$_6$ honeycomb and strongly buckled Fe2O$_6$ honeycomb layers along the $c$-axis. The nearest-neighbor interactions (J$_1$) between Fe cations are along the $c$ axis with the Fe1-Fe2 distance d$_1$=2.979(1)\AA~through superexchange paths Fe1-O1-Fe2 (91.8(2)$^o$) between the two edge-sharing FeO$_6$ as shown in Fig.\ref{fig:6}(c).

\subsection{MAGNETIC AND MAGNETODIELECTRIC COUPLING}
\begin{figure}
\centering
\includegraphics[width=1\linewidth]{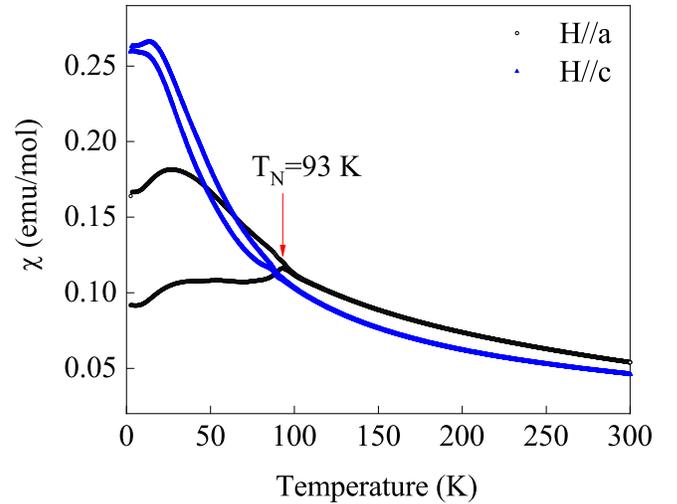}
\caption{(Color Online) Temperature dependence of the magnetic susceptibility of Fe$_4$Nb$_2$O$_9$ with a magnetic field H =0.2 T parallel to the $a$ and $c$ axes.}\label{fig:2}
\end{figure}

The magnetic susceptibility of a Fe$_4$Nb$_2$O$_9$ single crystal was measured with both zero field cooling (ZFC) and field cooling (FC) protocols on heating along the $a$ and $c$ axes. The evident difference between the magnetic susceptibility curves along the two directions indicates a strong easy-plane anisotropy. This may reflect that the magnetic moments are constrained in the $ab$ plane. Upon cooling, a single anomaly around T$_N$=93 K featured by a bifurcation between ZFC and FC curves was observed for both directions (Fig. \ref{fig:2}), which is indicative of a long-range antiferromagnetic transition, in accordance with the results previously reported on powder samples \citep{Maignan2018,Jana2019}. A small kink around 5 K is probably due to the presence of a small amount of impurity FeNb$_2$O$_6$ \citep{Heid1996}. 

\begin{figure}
\centering
\includegraphics[width=0.8\linewidth]{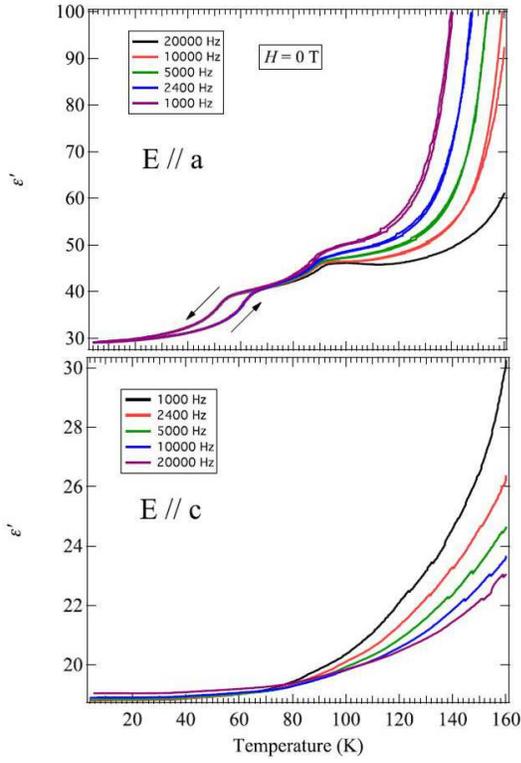}
\caption{(Color Online) The frequency-dependent dielectric constant with the electric field E parallel to the $a$- and $c$ axes, respectively.}\label{fig:3}
\end{figure}

To study the dielectric properties of Fe$_4$Nb$_2$O$_9$, the frequency-dependent dielectric constants ($\epsilon$) have been measured on the single crystal with an electric field along the $a$- or $c$ axis. The dielectric constant $\epsilon$'(T) along the $a$ axis is much larger than that along the $c$ axis, reflecting the presence of strong anisotropy of dielectric properties in  Fe$_4$Nb$_2$O$_9$. This is in fact consistent with the magnetic easy-plane anisotropy observed in the magnetic susceptibility curves. As shown in Fig. \ref{fig:3}, the anomalous change in the slope of dielectric constant is clearly discerned occurring at the N\'eel temperature for E$\parallel$a, but is invisible for E$\parallel$c. This indicates that the in-plane dielectric constant signals a much more sensitive response to the spin ordering than the magnetic susceptibility. Interestingly, below T$_N$, $\epsilon$'(T) decreases and shows another bump with the onset at T$_S$=70 K. We assign this anomaly to the crystal structural transition as observed in the previous work. \citep{Jana2019}.

\begin{figure}
\centering
\includegraphics[width=0.8\linewidth]{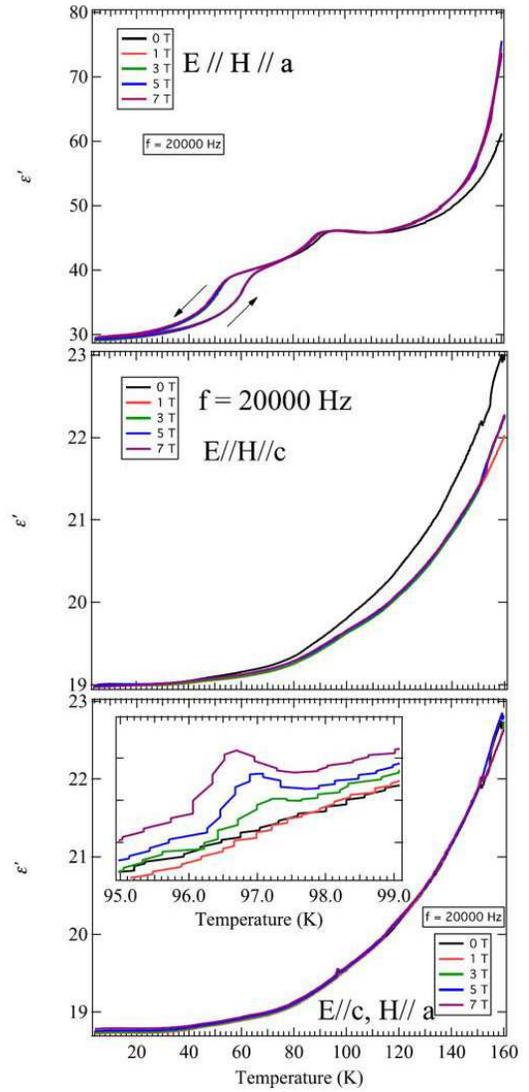}
\caption{(Color Online) Magnetic field dependence of the dielectric constant of Fe$_4$Nb$_2$O$_9$ with magnetic and electric fields applied to different configurations. Inset show the magnified view of the the magneto-electric coupling.}\label{fig:4}
\end{figure}

Magnetodielectric response has been considered to be a definitive and experimentally accessible method to recognize linear magnetoelectric materials.\citep{Mufti2011} This method can circumvent the experimental difficulties typically in the pyroelectric current measurements. We thus have evaluated the magnetoelectric coupling in Fe$_4$Nb$_2$O$_9$ single crystal by measuring the magnetodielectric response. Figure \ref{fig:4} shows the dielectric constant as a function of magnetic field with a fixed frequency f=20 KHz. Three configurations E$\parallel$H$\parallel$a, E$\parallel$H$\parallel$c and E$\parallel$c, H$\parallel$a have been used to measure the magnetodielectric response.
For the configuration E$\parallel$c, H$\parallel$a, a peak in the $\epsilon$'(T) curve was observed at T$_N$=93 K, which becomes stronger with increasing magnetic field. A small change at T$_N$ in the $\epsilon$'(T) was observed for E$\parallel$H$\parallel$a. However, no anomaly was observed up to 7 T for E$\parallel$H$\parallel$c. As we discussed in the following, these observations on the single crystal revealed by the magnetodielectric properties indicate that Fe$_4$Nb$_2$O$_9$ is a linear magnetoelectric material in intermediate magnetic fields \cite{Maignan2018}.

\begin{table} 
\caption{The refined structural parameters of Fe$_4$Nb$_2$O$_9$ with single crystal x-ray diffraction data at 40 K. Space group: $C2/c$, $a$=9.0861(6)\AA,  $b$=5.1970(4) \AA, $c$=14.234(1) \AA, $\beta$=91.294(3)$^o$. $\chi^2$=1.48. $B_{iso}$ has unit of 1/(8$\pi^2$)\AA$^2$.}\label{strlow}
\begin{tabular}{ccccccc}
\hline
\hline
atom & $x$& $y$ &  $z$  & $B_{iso}$ \\
\hline
Nb1 &  0.0010(3)& -0.0004(3)&  0.64284(6) & 0.87(2) \\
Fe1 & 0.8325(4) & 0.4994(4) &  0.6926(1) & 0.85(3) \\
Fe2 & 0.8337(4) & 0.5001(4) & 0.4855(1) & 0.86(3)\\
O1  & 0.845(3) & 0.821(2)& 0.0863(9) & 0.7(2)\\
O2  &  0.672(3) & 0.355(2)&  0.0858(8) & 1.5(2)\\
O3  &  0.986(3) & 0.328(2)&  0.0833(8) & 0.7(1)\\
O4  &  0 & 0.722(2)&  1/4 & 0.7(2)\\
O5  &  0.141(3) & 0.145(2)&  0.250(1) & 0.9(2)\\
\hline
\hline
\end{tabular}
\end{table}
\subsection{LOW TEMPERATURE CRYSTAL AND MAGNETIC STRUCTURES DETERMINATION}
\begin{figure}
\centering
\includegraphics[width=1\linewidth]{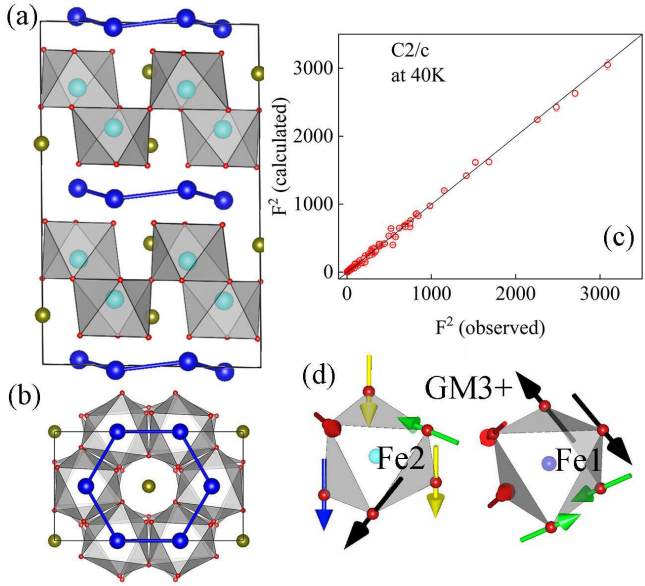}
\caption{(Color Online) (a-b) Crystal structure of Fe$_4$Nb$_2$O$_9$ projects along the $b$ and $c$ directions. (c) Observed structure factor squared values of diffraction peaks compared to those calculated by fitting the $C2/c$ monoclinic structural model at 40 K. (d) Visualization of the atomic displacements of the primary GM3+ irrep from the symmetry adapted mode analysis projected into the $ab$ plane.}\label{fig:5}
\end{figure}

To check the crystal structure at low temperatures, we have performed variable temperature single crystal x-ray (down to 40 K) and neutron diffraction (down to 5 K). For the neutron diffraction data,  rocking curve scans around Bragg peaks (0 3 0) and (1 1 3) at 5 K reveal obvious peak splittings, indicating a displacive transition. To solve the crystal structure at low temperatures, we have performed the group-subgroup analysis using tools at the Bilbao Crystallographic Server. In a sort of symmetry-related maximal subgroups of $P\bar{3}c1$, only subgroup $C2/c$ with the index of 3 is monoclinic. The structure was then refined in this space group using the single crystal x-ray diffraction data at 40 K. Figure \ref{fig:5}(c) shows the result of the refinement ($\chi^2$=1.45) as well as the crystal structure at 40 K. It appears that the crystal structure at 40 K is similar to the high temperature one. O1 and O2 in the $P\bar{3}c1$ structure were split into five atomic positions in the low temperature structure, demonstrating the largest atomic displacements. To illustrate the O atomic displacements compared to the parent structure ($P\bar{3}c1$) we have conducted the symmetry mode analysis by considering the two space groups using AMPLIMODES. \citep{bilbao} The decomposition of the low temperature structure in respect of the symmetrized displacive modes of the parent $P\bar{3}c1$ structure yields a set of symmetry-adopted distorted modes that are associated with two irreducible representation(irreps) GM1+ and GM3+. Through the analysis, we found that GM3+ transforms all the primary modes. We have drawn the displacements corresponding to each symmetry mode of all O atoms (which form the distorted octahedra FeO$_6$) in Fig. \ref{fig:5}(d). It is evident that the largest distortions occur mainly in the $ab$ plane.          

Although the magnetic structure of Fe$_4$Nb$_2$O$_9$ has been previously reported by studying a powder sample using neutron diffraction, the exact magnetic symmetry is still unclear as powder neutron diffraction can not distinguish the magnetic space groups $C2'/c$ and $C2/c'$.\citep{Jana2019} We have carried out single crystal neutron diffraction on Fe$_4$Nb$_2$O$_9$ to clearly sort out this problem. Even though Fe$_4$Nb$_2$O$_9$ undergoes a crystal structure phase transition at 70 K with splittings of some specific Bragg peaks, no anomaly was observed in the magnetic reflections, indicating that the magnetic symmetry is preserved upon cooling. Therefore, hereafter, we will describe the magnetic structure in the parent structure form. At 5 K, all observed magnetic reflections were well indexed by a vector \textbf{k}=\textbf{0}, consistent with that found in the powder sample. \citep{Jana2019}.

\begin{figure}
\centering
\includegraphics[width=1\linewidth]{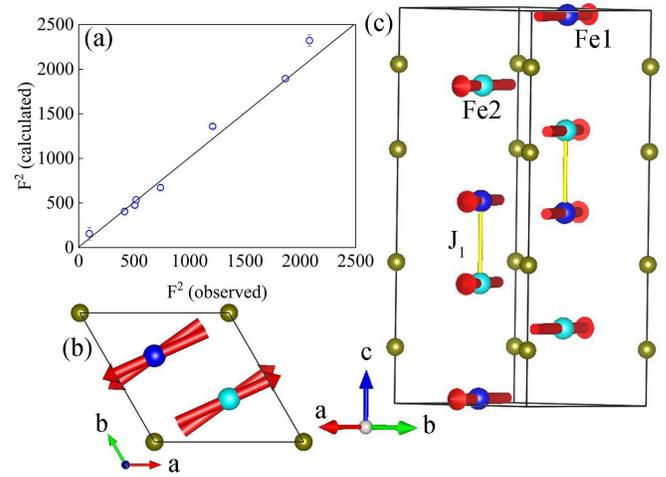}
\caption{(Color Online) (a) Observed structure factor squared values
of all three families of diffraction peaks at 5 K compared to those calculated by fitting the magnetic structural model (b-c) The ground state magnetic structure of Fe$_4$Nb$_2$O$_9$ in the hexagonal lattice setting.}\label{fig:6}
\end{figure}

A set of magnetic subgroups that are compatible with the given space group and the propagation vector was obtained through the symmetry analysis using Bilbao Crystallographic Server (Magnetic Symmetry and Applications \citep{bilbao}) software. The magnetic structures containing trigonal symmetry imply a magnetic configuration with the moments along the $c$-axis, inconsistent with our experimentally measured magnetic reflections such as (0 0 4) and thermodynamic measurements. Thus, we have to lower the symmetry in the magnetic subgroup hierarchy. This gives four monoclinic magnetic subgroups $C2'/c'$, $C2/c'$, $C2'/c$ and $C2/c$. Without refining our neutron data, one can rule out the subgroups $C2'/c'$ and $C2/c$ by applying Neumann’s principle to the magnetoelectric effect. Basically, both of them do not allow the linear magneoelectric effect. We have tested the remaining magnetic models using the single crystal neutron diffraction data. The model corresponding to the subgroup $C2/c'$ yields a satisfactory fit to our data ($R_f$=4.43\% and $\chi^2$=5.13), as shown in Fig. \ref{fig:6}(a). In fact, another reasoning of precluding the other candidate $C2'/c$ is based on the magnetoelectric coupling experimentally observed in Fe$_4$Nb$_2$O$_9$ since the $C2'/c$ subgroup allows only the off-diagonal terms. Therefore, we can conclude that the magnetic structure of Fe$_4$Nb$_2$O$_9$  is described by the magnetic space group $C2/c'$. Fig.\ref{fig:4} shows the magnetic structure of Fe$_4$Nb$_2$O$_9$, which is manifested as antiferromagnetically coupled ferromagnetic chains along the $c$-axis with all spins confined into the $ab$ plane. In particular, the nearest neighboring Fe atoms, i.e., Fe1 and Fe2 order ferromagnetically. This pair forms a small canting angle 5.81$^{\circ}$ with its adjacent Fe1-Fe2 pair in the chain. This canting angle is roughly two times smaller that that in Co$_4$Nb$_2$O$_9$ \citep{Ding2020}. The refined magnetic moment at 5 K is 3.52(4) $\mu_B$ for both the Fe1 and Fe2 sites. This value is slightly smaller than the theoretical ordered spin-only value 4 $\mu_B$ for Fe$^{2+}$ with a high spin state. Due to the existence of the trigonal lattice symmetry, three magnetic domains were considered and their populations were set to be equal during the refinement. 

\section{DISCUSSION}

We have established that Fe$_4$Nb$_2$O$_9$ undergoes an antiferromagnetic phase transition at 93 K, followed by a crystal structure transition at 70 K. By combining x-ray and neutron diffraction, we have solved the crystal structure and magnetic symmetry at low temperatures. The magnetic structure is featured by the ferromagnetic chain with a small canting between each adjacent Fe1-Fe2 pair. This indicates that the dominant exchange interaction J$_1$ is ferromagnetic. Owing to the presence of the antiferromagnetic exchange interactions (J$_2$) between Fe1 atoms in the slightly buckled honeycomb in the $ab$ plane, the ferromagnetic chains are coupled antiferromagnetically. The strong competition between the two sorts of exchange interactions may be responsible for the small tilting angle between each ferromagnetic Fe1-Fe2 pair. 
   
Let us first discuss the anomalies observed in $\epsilon$'(T) without magnetic fields. The bump at 93 K is similar to the dielectric properties observed in MnO and MnF$_2$ where the anomalies were explained by the presence of exchange striction due to the spin-lattice coupling.  \citep{Seehra1981,Seehra1984} In our case, the formation of the long range magnetic ordering with all spins in the $ab$ plane could be a driving force for the occurrence of exchange striction, which in turn induces the anomaly only in the dielectric constant with the E$\parallel$a. The more apparent anomaly observed in the dielectric constant curve with the E$\parallel$a at T$_S$=70 K is due to the crystal structural transition which is predominantly triggered by the atomic displacements of O atoms, as shown in Fig. \ref{fig:5}(d).  

When applying the magnetic symmetry $C2/c'$ of Fe$_4$Nb$_2$O$_9$ to the magnetoelectric coupling tensor, we have in principle five non-zero terms: $\alpha_{xx}$, $\alpha_{xz}$, $\alpha_{yy}$, $\alpha_{zx}$, $\alpha_{zz}$. Our temperature and magnetic field dependence of the dielectric constant measurements along the different directions of the crystal have shown that the strongest term is $\alpha_{zx}$ and the term $\alpha_{xx}$ is small while the term $\alpha_{zz}$ is null. 
The off-diagonal component $\alpha_{zx}$ with a large value in principle allows the occurrence of the ferrotoroidal order in Fe$_4$Nb$_2$O$_9$. The experimental observation of the relevant ferrotoroidal domains through an essential magnetoelectric annealing process is greatly desired.

\section{CONCLUSION}
In summary, we have studied the magnetoelectric coupling and magnetic structure of Fe$_4$Nb$_2$O$_9$ single crystal. The magnetic susceptibility and single crystal neutron diffraction showed an antiferromagnetic transition at 93 K described by the $C2/c'$ magnetic space group with two antiparallel  ferromagnetic chains totally confined in the $ab$ plane. Along the $c$ axis, a small but pivotal tilting angle between each adjoining Fe1-Fe2 atoms was found. This magnetic symmetry essentially triggers the occurrence of exchange striction that explains the anomaly in the dielectric constant in the $a$ direction, and allows the linear ME effect as observed experimentally by the magnetodielectric response measurements. 
Our work also suggests that manipulation of the magnetoelectric coupling through an external magnetic field is practicable since the magnetic configuration in honeycomb Fe$_4$Nb$_2$O$_9$ makes the continuous rotation of the moments via a magnetic field robust.
  
\section{EXPERIMENTAL METHODS}
Single crystals of Fe$_4$Nb$_2$O$_9$ were grown by the traveling-solvent floating-zone (TSFZ) technique. The feed and seed rods for the crystal growth were prepared by solid state reaction. Appropriate mixtures of Fe$_2$O$_3$, Fe, and Nb$_2$O$_5$ were ground together and pressed into 6 mm diameter $\times$ 60 mm rods under 400 atm hydrostatic pressure and then calcined in Argon at 1100 Celsius degree for 24 h. The crystal growth was carried out in argon in an IR-heated image furnace (NEC) equipped with two halogen lamps and double ellipsoidal mirrors with feed and seed rods rotating in opposite directions at 25 rpm during crystal growth at a rate of 4 mm/h.

Single-crystal x-ray diffraction data were collected down to 40 K using a Rigaku XtaLAB PRO diffractometer with the graphite monochromated Mo $K\alpha$ radiation ($\lambda$ = 0.71073 \AA) equipped with a HyPix-6000HE detector and an Oxford N-HeliX cryocooler. Peak indexing and integration were done using the Rigaku Oxford Diffraction CrysAlisPro software \citep{rigaku}. An empirical absorption correction was applied using the SCALE3 ABSPACK algorithm as implemented in CrysAlisPro. \citep{higashi2000}. Structure refinement was done using FULLPROF Suite \cite{fullprof}.

The dc magnetization curves were obtained using a vibrating sample magnetometer (VSM). For the dielectric constant measurements, single crystals were used whose orientations were determined by Laue diffraction. Two single crystalline samples were polished to achieve two parallel flat surfaces perpendicular to the $a$ and the $c$ axes. An Andeen-Hagerling AH-2700A commercial capacitance bridge was used to measure the capacitance, which was converted to dielectric constant. This measurement employed electric fields of 21 and 14.6 kV/m for E$\parallel$a and E$\parallel$c configuration, respectively.

Single-crystal neutron diffraction was performed at the HB-3A Four-Circle Diffractometer (FCD) equipped with a 2D detector at the High Flux Isotope Reactor (HFIR) at Oak Ridge National Laboratory (ORNL). Neutron wavelengths of 1.003~\AA~ (neutron energy 81 meV) were used with a bent perfect Si-331 monochromator \cite{hb3a}. The nuclear and magnetic structure refinements were performed with the FULLPROF Suite \cite{fullprof}.

\section{DATA AVAILABILITY}
All relevant data are available from the corresponding author upon request.

\begin{acknowledgments}
The research at Oak Ridge National Laboratory
(ORNL) was supported by the U.S. Department of Energy
(DOE), Office of Science, Office of Basic Energy Sciences,
Early Career Research Program Award KC0402010, under
Contract DE-AC05-00OR22725 and the U.S. DOE, Office of
Science User Facility operated by the ORNL. The work at University of Tennessee was supported by DOE under award DE-SC-0020254. A portion of this work was performed at the National High Magnetic Field Laboratory, supported by the National Science Foundation Cooperative Agreement No. DMR-1644779 and the State of Florida.  
The US Government retains, and
the publisher, by accepting the article for publication, acknowledges
that the US Government retains a nonexclusive,
paid-up, irrevocable, worldwide license to publish or reproduce
the published form of this manuscript, or allow others to do so, for US Government purposes. The Department of
Energy will provide public access to these results of federally
sponsored research in accordance with the DOE Public Access
Plan.\citep{DOE}
\end{acknowledgments}

\section{AUTHOR CONTRIBUTIONS}
R. S. and H. Z. grew the crystals. M. L. and E. S. C. characterized the crystals. Y. W. and B. C. C. did the single crystal x-ray diffraction. L. D. and H. C. carried out the neutron diffraction, analyzed the data and wrote the manuscript with inputs from all co-authors.

\section{Competing Interests}
The Authors declare no Competing Financial or Non-Financial Interests


\begin{thebibliography}{99}

\bibitem{schmid1994} H. Schmid, Ferroelectrics 162, 317 (1994).
\bibitem{rivera2009} J.-P. Rivera, Eur. Phys. J. B 71, 299–313 (2009)  % A short review of the magnetoelectric effect and related experimental techniques on single phase (multi-) ferroics
\bibitem{fiebig2005} M. Fiebig, J. Phys. D 38, R123 (2005). % Revival of the magnetoelectric effect
\bibitem{tokura2014} Y. Tokura, S. Seki, and N. Nagaosa, Rep. Prog. Phys. 77, 076501 (2014).
\bibitem{spaldin2008} N. A. Spaldin, M. Fiebig, and M. Mostovoy, J. Phys.: Condens. Matter 20, 434203 (2008).
\bibitem{schmid2008} H. Schmid, J. Phys.: Condens. Matter 20, 434201 (2008). 
\bibitem{wang2003} J. Wang, J. B. Neaton, H. Zheng1, V. Nagarajan, S. B. Ogale, B. Liu, D. Viehland, V. Vaithyanathan, D. G. Schlom, U. V. Waghmare, N. A. Spaldin, K. M. Rabe, M. Wuttig, R. Ramesh, Science 299, 1719 (2003).  %Epitaxial BiFeO3 Multiferroic Thin Film Heterostructures 
\bibitem{spaldin2005} N. A. Spaldin, M. Fiebig: The renaissance of magnetoelectric multiferroics, Science 309, 391 (2005).  %The Renaissance of Magnetoelectric Multiferroics
\bibitem{eerenstein2006} W. Eerenstein, N. D. Mathur, and J. F. Scott, Nature (London) 442, 759 (2006). %Multiferroic and magnetoelectric materials
\bibitem{cheong2007} S.-W. Cheong and M. Mostovoy, Nat. Mater. 6, 13 (2007). %Multiferroics: a magnetic twist for ferroelectricity
\bibitem{ding2016} L. Ding, C. V. Colin, C. Darie, J. Robert, F. Gay, and P. Bordet, Phys. Rev. B 93, 064423 (2016).
\bibitem{Fischer1972} E. Fischer, G. Gorodetsky, and R. M. Hornreich, Solid State Commu. 10, 1127 (1972).
%first ME in MnNbO and CoNbO
\bibitem{Fang2015} Y. Fang, W. P. Zhou, S. M. Yan, R. Bai, Z. H. Qian, Q. Y. Xu, D. H. Wang, and Y. W. Du, J. Appl. Phys. 117, 17B712 (2015).
\bibitem{fang2014} Y. Fang et al., Sci. Rep. 4 3860 (2014). %Large magnetoelectric coupling in Co4Nb2O9
\bibitem{khanh2016} N. D. Khanh, N. Abe, H. Sagayama, A. Nakao, T. Hanashima, R. Kiyanagi, Y. Tokunaga, and T. Arima, Phys. Rev. B 93, 075117 (2016).
\bibitem{bertaut1961} E.F. Bertaut, L.Corliss, F. Forrat, R. Aleonard, and R. Pauthenet, J. Phys. Chem. Solids 21, 234 (1961). %first neutron diffraction on MnNbO and CoNbO
\bibitem{Dzyaloshinskii1959} I.E. Dzyaloshinskii, Zh. Exp. Teor. Fiz. 37, 881 (1959) [Soviet Phys. JETP 10, 628 (1960)]
\bibitem{astrov1960} D.N. Astrov, Zh. Exp. Teor. Fiz. 38, 984 (1960) [Soviet Phys. JETP 11, 708 (1960)]
\bibitem{kimura2013} A. Iyama and T. Kimura, Phys. Rev. B 87, 180408(R) (2013)
\bibitem{fiebig1994} M. Fiebig, D. Frohlich, B. B. Krichevtsov, and R. V. Pisarev, Phys. Rev. Lett. 73, 2127 (1994).
\bibitem{mcgurie1956} T. R. McGuire, E. J. Scott, and F. H. Grannis, Phys. Rev. 102, 1000 (1956).
\bibitem{deng2018} G. Deng, Y. Cao, W. Ren, S. Cao, A. J. Studer, N. Gauthier, M. Kenzelmann, G. Davidson, K. C. Rule, J. S. Gardner, P. Imperia, C. Ulrich, and G. J. McIntyre, Phys. Rev. B 97, 085154 (2018).
\bibitem{khanh2017} N. D. Khanh, N. Abe, S. Kimura, Y. Tokunaga, and T. Arima, Phys. Rev. B 96, 094434 (2017).
\bibitem{Ding2020} L. Ding et al. arXiv preprint arXiv:2002.10599 (2020).
\bibitem{Maignan2018} A. Maignan, and C. Martin, Phys. Rev. B 97, 161106(R)(2018)
\bibitem{Jana2019} R. Jana, et al. Phys. Rev. B 100, 094109 (2019)
\bibitem{rigaku} Rigaku, (2005) CrystalClear. Rigaku Corporation, Tokyo, Japan.
\bibitem{higashi2000} T. Higashi, ABSCOR (2000). Rigaku Corporation, Tokyo, Japan.
\bibitem{fullprof} J. Rodriguez-Carvajal, Physica B 192 55 (1993).
\bibitem{Heid1996} C. Heid, H. Weitzel, F. Bourdarot, R. Calemczuk, T. Vogt, and H. Fuess, J. Phys.: Condens. Matter 8, 10609 (1996).
\bibitem{Mufti2011} N. Mufti, G. R. Blake, M. Mostovoy, S. Riyadi, A. A. Nugroho, and T. T. M. Palstra, Phys. Rev. B 83, 104416 (2011).
\bibitem{Seehra1981} M. S. Seehra, and R. E. Helmick, Phys. Rev. B 24, 5098 (1981).
\bibitem{Seehra1984} M. S. Seehra, and R. E. Helmick, J. Appl. Phys. 55, 2330 (1984). 
\bibitem{hb3a} B. C. Chakoumakos, H. Cao, F. Ye, A. D. Stoica, M. Popovici, M. Sundaram, W. Zhou, J. S. Hicks, G. W. Lynn, and R. A. Riedel, J. Applied Cryst. 44, 655 (2011).
\bibitem{bilbao} J. M. Perez-Mato, S. V. Gallego, E. S. Tasci, L. Elcoro, G. de la Flor, and M. I. Aroyo, Annu. Rev. Mater. Res. 45, 217 (2015).

\bibitem{DOE}Https://www.energy.gov/downloads/doe-public-access-plan.


 


\end{thebibliography}
\end{document}